\documentclass[runningheads]{llncs}

\usepackage{amsmath}
\usepackage{amsfonts}
\usepackage{amssymb}
\usepackage{graphicx}
\usepackage{todonotes}
\usepackage{booktabs}
\usepackage[switch]{lineno}
\usepackage{longtable}
\usepackage[colorlinks]{hyperref}
\usepackage{placeins}
\usepackage{subcaption}
\usepackage{bm}

\newcommand{\ignore}[1]{}

\newcommand{\winners}{W}

\newcommand{\cand}{\mathcal{C}}

\newcommand{\ballots}{\mathcal{B}}
\newcommand{\quota}{\mathcal{Q}}
\newcommand{\seats}{N}
\newcommand{\election}{E}

\newcommand{\AG}{\textsf{AG}}
\newcommand{\IQ}{\textsf{IQ}}
\newcommand{\UT}{\textsf{UT}}
\newcommand{\LT}{\textsf{LT}^*}
\newcommand{\CNEB}{\textsf{NL}}
\newcommand{\CNEBS}{\textsf{NL}^*}
\DeclareMathOperator{\first}{first}
\newcommand{\proj}{\sigma}
\newcommand{\AGS}{\textsf{AG}^*}

\title{RLAs for 2-Seat STV Elections: Revisited\thanks{To appear in the 9th
Workshop on Advances in Secure Electronic Voting (Voting'24), on 8 March 2024.
This work was partially supported by
the Australian Research Council: Discovery Project DP220101012, OPTIMA ITTC
IC200100009.}}

\author{
Michelle Blom     \inst{1}   \orcidID{0000-0002-0459-9917}  \and
Peter J. Stuckey  \inst{2}   \orcidID{0000-0003-2186-0459}  \and
Vanessa Teague    \inst{3}   \orcidID{0000-0003-2648-2565}  \and
Damjan Vukcevic   \inst{4} \orcidID{0000-0001-7780-9586}}

\authorrunning{Blom, Stuckey, Teague, Vukcevic}

\institute{
School of Computing and Information Systems, University of Melbourne,\\
Parkville, Australia \\
\email{michelle.blom@unimelb.edu.au}
\and
Department of Data Science and AI, Monash University, Clayton, Australia
\and
Thinking Cybersecurity Pty.\ Ltd., Melbourne, Australia
\and
Department of Econometrics and Business Statistics, Monash University, \\
Clayton, Australia}

\begin{document}

\maketitle

\begin{abstract}
    Single Transferable Vote (STV) elections are a principled approach to electing multiple candidates in a single election. Each ballot has a starting value of 1, and a candidate is elected if they gather a total vote value more than a defined quota. 
    Votes over the quota have their value reduced by a \emph{transfer value} so as to remove the quota, and are passed to the next candidate on the ballot. 
    Risk-limiting audits (RLAs) are a statistically sound approach to election auditing which guarantees that failure to detect an error in the result is bounded by a limit.  A first approach to RLAs for 2-seat STV elections has been defined. In this paper we show how we can improve this approach by reasoning about lower bounds on transfer values, and how we can extend the approach to \emph{partially} audit an election, if the method does not support a full audit.
\end{abstract}

% ---------------------------------------------------------------------------

\section{Introduction}

Single Transferable Vote (STV) elections are widely used around the world for multiple candidate contests. Risk-limiting audits (RLAs) are very complex for STV elections. Prior work has demonstrated that RLAs for some 2-seat STV elections, where at least one candidate has more than quota's worth of votes on their first preferences, are possible \cite{blom2022stvrlasSHORT}. This existing work proposed two approaches for undertaking RLAs for 2-seat STV. The first tackles the case where this \textit{first-round winner criterion} is satisfied, while the second presents a general method that applies when it does not. The latter method was generally not successful in forming an audit to verify the correctness of both winners. 

This paper presents an improved method for 2-seat STV RLAs where the contest satisfies the first-round winner criterion. This new method is able to form audits for a greater number of contests, and reduces the expected sample sizes needed for these audits by 15 to 19\% across the contests considered in our evaluation. The original method introduced assertions for (i) verifying that the first winner achieved a quota on their first preferences, (ii) verifying an upper bound on the transfer value of this first winner, and (iii) using this upper bound, verifying that the second winner could not have possibly lost to any of the reported losers. We improve this method by introducing a new assertion that verifies a non-trivial \textit{lower bound} on this first winner's transfer value, and using both bounds to fine tune the assertions formed in (iii). We additionally show how contests that perform a preliminary batch-elimination prior to electing any candidates can be audited using this scheme.

A full RLA, verifying the correctness of both winners, may not possible for a given 2-seat contest. It may be desirable to perform some kind of audit to verify some aspects of the reported outcome. We show how the `general' method can be re-framed as a five-stage process that forms a \textit{partial} RLA. This process aims to establish (i) a subset of candidates as definite losers, and (ii) a subset of candidates as definite winners. The remaining candidates are \textit{possible} winners. The first three stages of this revised general method are drawn from the work of \cite{blom2022stvrlasSHORT}. In this paper we add a fourth stage that looks for opportunities to reduce the expected sample size of the partial audit. The final stage summarises what aspects of the reported outcome are verified by the audit, and which are not. 

The remainder of this paper is structured as follows.
\autoref{sec:Preliminaries} describes the variant of STV that we consider, and
assertion-based RLAs.
Three sections follow that consider different election scenarios and how to
audit them:
\autoref{sec:BatchFirst} covers auditing of batch elimination,
\autoref{sec:FWC} covers the improved \textit{first-round winner} method and an
evaluation against the existing approach, and
\autoref{sec:GeneralMethod} shows how we can partially audit elections where no
candidate has a quota on first preferences.
We conclude in \autoref{sec:conc}.

% ---------------------------------------------------------------------------

\section{Preliminaries}\label{sec:Preliminaries}

We consider a variant of STV, modelled on how STV is typically implemented in the United States.  We describe this variant in \autoref{sec:USSTV}. 

We define an STV election as per \autoref{def:STV}. We define a ballot $b$ as a sequence of candidates $\pi$, listed in order of preference (most popular first), without
duplicates but without necessarily including all candidates. We use list
notation (e.g., $\pi = [c_1,c_2,c_3,c_4]$). The notation $\first(\pi) = \pi(1)$
denotes the first item (candidate) in  sequence $\pi$. 

\begin{definition}[STV Election]\label{def:STV}
An STV election $\election$ is a tuple $\election = (\cand, \ballots, \quota,
\seats)$ where $\cand$ is a set of candidates,  $\ballots$ the multiset of
ballots cast\footnote{A multiset allows for the
inclusion of duplicate items.}, $\quota$ the election quota (the number of votes a candidate must
attain to win a seat---usually the Droop quota---\autoref{eqn:Droop}), and
$\seats$ the number of seats to be filled.
\begin{equation}
\quota = \left\lfloor \frac{|\ballots|}{\seats + 1} \right\rfloor + 1
\label{eqn:Droop}
\end{equation}
\end{definition}

\begin{definition}{\textbf{Projection} $\mathbf{\proj_\mathcal{S}(\pi)}$}
We define the projection of a sequence $\pi$ onto a set $\mathcal{S}$ as the
largest subsequence of $\pi$ that contains only elements of $\mathcal{S}$. (The
elements keep their relative order in $\pi$.) For example:
$\proj_{\{c_2,c_3\}}([c_1,c_2,c_4,c_3]) = [c_2,c_3]$ 
 and $\proj_{\{c_2,c_3,c_4,c_5\}}([c_6,c_4,c_7,c_2,c_1]) = [c_4,c_2].$
\label{def:Projection}
\end{definition}

The tabulation of STV elections proceeds in rounds (see \autoref{sec:USSTV}). Initially, all candidates are awarded the ballots on which they are the first ranked candidate. We call a candidate $c$'s tally at this stage their \textit{first-preference tally}, denoted $t_{c,1}$. We use $t_{c,r}$ to denote a candidate's tally at the start of round $r$ of tabulation. In the election of \autoref{tab:EGSTV1}, candidates $c_1$ to $c_5$ have first-preference tallies of 9001, 3000, 5000, 3950, and 50 votes, respectively. 

\begin{table}[t]
% FP tallies: c1 (9001); c2 (3000); c3 (5000); c4 (3950); c5 (50)
\centering
\begin{subtable}{.5\columnwidth}
\begin{tabular}{lrclr}
\toprule
Ranking & Count & & Ranking & Count\\
\midrule
{}[$c_1$, $c_3$]        &  8,001  & $\quad$ & {}[$c_3$, $c_4$]        &  5,000 \\
{}[$c_1$]               &  1,000  &  $\quad$ & {}[$c_4$, $c_1$, $c_2$]        &  3,950 \\
{}[$c_2$, $c_3$, $c_4$] &  3,000  &  $\quad$ & {}[$c_5$, $c_2$]               &    50\\
\midrule
Total                   &  & & & 21,001 \\
\bottomrule
\end{tabular}
\end{subtable}
\begin{subtable}{0.2\columnwidth}
\begin{tabular}{ll}
$\seats = 2$  & \\
$\quota = 7,001$ & \\
$t_{c_1,1} = 9,001$ & $\quad t_{c_4,1} = 3,950$ \\
$t_{c_2,1} = 3,000$ & $\quad t_{c_5,1} = 50$ \\
$t_{c_3,1} = 5,000$ & \\
\end{tabular}
\end{subtable}
\caption{ An STV election, stating the number of
ballots cast with each listed ranking over candidates $c_1$ to $c_5$. The quota, and first-preference tallies are listed.}
\label{tab:EGSTV1}
\end{table}

\subsection{`US' style STV}\label{sec:USSTV}

Each ballot cast in the election starts with a value of 1. In any given round of tabulation, if no candidate's tally is equal to or above the election's quota, the candidate with the smallest tally is eliminated. All the ballots in the eliminated candidate's tally are redistributed to the next most preferred \textit{eligible} candidate on the ballot. These ballots are transferred at their current value. At any stage where ballots are distributed from one candidate to another, the only candidates that are eligible to receive votes are those that have not yet been eliminated or elected to a seat, and who have less than a quota's worth of votes in their tally. 

Eliminations proceed as described above until at least one candidate's tally
equals or exceeds the election's quota. At this stage, these candidates are
elected to a seat. These candidates will be elected to a seat in order of their
\textit{surplus}. For each such candidate, we define their \textit{surplus} as
the difference between their current tally and the quota. The ballots sitting
in the tally pile of this candidate are \textit{reweighted} and distributed to
the next most preferred eligible candidate on the ballot. For a  candidate $c$,
elected to a seat in round $r$ of tabulation, we define the \textit{transfer
value} $\tau_c$  used to reweight the ballots in their tally as shown in
\autoref{eqn:TransferValueUS}, where $t_{c,r}$ denotes the tally of $c$ at the
start of round $r$.

\begin{equation}
    \tau_c = \frac{t_{c,r} - \quota}{t_{c,r}}
    \label{eqn:TransferValueUS}
\end{equation}

\noindent For a ballot $b \in \ballots$, whose current value is $v_{b,r}$, its new value when it leaves the tally pile of candidate $c$ upon their election to a seat becomes $\tau_c \, v_{b,r}$.

Tabulation proceeds by seating candidates whose tally reaches or exceeds a quota and distributing their votes to eligible candidates, and eliminating candidates when no candidate has a quota. This process continues until either all seats have been filled, or the number of candidates still standing equals the number of seats left to be filled. These remaining candidates are then elected.

Consider the election in \autoref{tab:EGSTV1}. The quota is 7001 votes. Candidate $c_1$ has a quota on  first preferences, and is elected to the first seat. Their transfer value is $\tau = (9001 - 7001)/9001 = 0.222$. The 8,001 {}[$c_1$, $c_3$] ballots are given to candidate $c_3$, adding 1,776.222 votes to $c_3$'s tally. The 1,000 {}[$c_1$] ballots become \textit{exhausted}. Candidate $c_3$ now has 6,776.222 votes. As no candidate has a quota's worth of votes, the candidate with the smallest tally is eliminated. Here,  this is candidate $c_5$ on 50 votes. These 50 ballots are given to $c_2$ at their current value of 1. Candidate $c_2$ now has 3,050 votes. Still no candidate has a quota's worth of votes. Candidate $c_2$ is eliminated next. The 50  {}[$c_5$, $c_2$] ballots become exhausted, and the 3,000 {}[$c_2$, $c_3$, $c_4$] ballots are given to $c_3$, who now has 9,776.222 votes. Candidate $c_3$ has achieved a quota, and is elected to the second seat.

\textbf{Batch elimination} We also consider a variation of the above process in which  a  batch elimination step is first performed. We first determine if there are any candidates for which there is no mathematical possibility for them to win. For each candidate, we compute the number of ballots on which they are ranked, and compare this tally to the tally of the $N$ candidates with the highest first-preference tally. Consider the election in \autoref{tab:EGSTV1}. The two candidates with the highest first-preference tallies are $c_1$ on 8,001 votes, and $c_3$ on 5,000 votes. Candidate $c_5$ is ranked on 50 ballots. Candidates $c_2$ and $c_4$ are ranked on 7,000 and 11,950 ballots, respectively. There is no possibility for $c_5$ to win, so they are eliminated in the first round, and their 50 {}[$c_5$, $c_2$] ballots are given to $c_2$. 

Tabulation then proceeds as described above. Candidate $c_1$ is elected, giving $c_3$ 1,776.2 votes and a tally of 6,776.2. Candidate $c_2$ is eliminated, giving 3,000 votes to $c_3$. Candidate $c_3$ is elected to the second seat on a tally of 9,776.2.

\subsection{Assertion-based Approaches to Risk-Limiting Audits}

SHANGRLA~\cite{shangrlaSHORT} provides a general framework for RLAs, using \emph{assertions} as `building blocks'.  An \emph{assertion} is a statement about the full set of ballots in an election.  These are typically expressed as an inequality about some property that would be consistent with a particular election outcome.  An example of an assertion is ``Alice received more votes than Bob''.
In the SHANGRLA framework, we need to design a set of assertions such that, if they are all true, they imply that the reported winner really won the election.  To conduct an audit, we statistically test each assertion using general statistical methods that form part of the framework.
Assertions need to have a specific mathematical form to fit into the framework.  In general, any linear combination of tallies (counts of different types of ballots) can be converted into a SHANGRLA assertion \cite{blom2021assertionSHORT}.  All of the assertions we develop in this paper are of this form.

% ---------------------------------------------------------------------------

\section{Context: Batch Elimination First}\label{sec:BatchFirst}

We first consider how we can verify, in an RLA, that the candidates eliminated as part of an initial batch elimination did indeed have no mathematical possibility of winning. To do so, we use the existing $\AG$ assertion of Blom~\emph{et al}~\cite{blom2022stvrlasSHORT}. 

\begin{description}
\item[$\AG(w,l)$]
Verifies that candidate $w$ always has a higher tally than $l$ by showing that $w$'s first-preference tally is higher than the maximum tally $l$ could achieve while $w$ is still standing: $t_{w,1} > | \{ b : b \in \ballots, \first\left(\proj_{\{w,l\}}(b)\right) = l \}|$.    
\end{description}

For a US-STV election $\election = (\cand, \ballots, \quota,
\seats)$, let $Top = \{c_1,\ldots,c_\seats\} \subset \cand$ denote the $\seats$  candidates with the highest first-preference tallies, and $Batch \subset \cand$ the set of batch eliminated candidates. We verify that candidate $c \in Batch$  cannot win by showing that $\AG(c_i, c)$ for all $c_i \in Top$. 

\begin{example}
\label{ex:2021BoEBatchElim}
    We consider the 2021 Board of Estimates and Taxation election in Minneapolis, Minnesota. 
This two-seat STV election involved four candidates -- S. Brandt, S. Pree-Stinson, P. Salica, and K. Nikiforakis -- and a number of undeclared write-ins (UWIs). The first-preference tallies  were 42672 votes for S. Brandt, 25597 votes for S. Pree-Stinson, 20786 votes for P. Salica, 5815 votes for K. Nikiforakis and 755 votes for UWIs. The quota or election threshold was 31876, and out of 145337 ballots, 49712 of these were invalid. The UWIs and K. Nikiforakis were eliminated in the first round with too few mentions to have a mathematical possibility of winning. S. Brandt was then elected, P. Salica was eliminated, leaving S. Pree-Stinson as the second winner.
We verify this batch elimination in an RLA with the assertions shown in
\autoref{tab:2021BoEBatchElim}, alongside the expected number of ballots
required to audit them.\footnote{For all sample size estimations, we assume
a risk limit of 10\%, an error rate of 2 overstatements per 1000 ballots,
and the ALPHA risk function of SHANGRLA \cite{shangrlaSHORT}.}
\qed
\end{example}
\begin{table}[t]
\caption{Assertions verifying the batch elimination of the UWIs and  Nikiforakis in the 2021 BoE election in Minneapolis, Minnesota, and their sample sizes.  }
\centering
\begin{tabular}{lr}
\hline
Assertion & Sample Size \\
\hline
    \AG(S. Brandt, UWIs) &  20  \\
    \AG(S. Pree-Stinson, UWIs) &  35 \\
    \AG(S. Brandt, K. Nikiforakis) &  27  \\
    \AG(S. Pree-Stinson, K. Nikiforakis) &  69  \\
\hline
Total cost: & 69 \\
\hline
\end{tabular}
\label{tab:2021BoEBatchElim}
\end{table}

\begin{example}
Batch Elimination First can change the result of an election.
Consider the two-seat STV with candidates $\{w, a, b, c_1, c_2, c_3, c_4, c_5 \}$
and ballots $[w]: 15001$, $[a]:6875$, $[b]:3125$, 
$[c_1,w,b]:1000$,
$[c_2,w,b]:1000$,
$[c_3,w,b]:1000$,
$[c_4,w,b]:1000$,
$[c_5,w,b]:1000$. The Droop Quota is 10001.
Without batch elimination we give a seat to $w$, then all the votes for $w$ exhaust; then each of the $c_i$ are eliminated leaving tallies $a: 6825$
and $b: 8125$, and finally $b$ is seated. 

With a batch elimination all of $c_1, \ldots, c_5$ are eliminated first, none of them has enough mentions to beat $w$, $a$, or $b$. Then $w$ gets a seat with 20001 votes in its tally and each of its surviving votes are transferred at value 0.5. The tallies for $a$ and $b$ are then $a: 6875$ and $b: 5625$, so finally $a$ is seated.
\qed
\end{example}

% ---------------------------------------------------------------------------

\section{Context: First Round Winner}\label{sec:FWC}

Let $\election = (\cand, \ballots, \quota, \seats = 2)$ denote a US-STV election with winners $w_1,w_2 \in \winners$ in which candidate $w_1$ is elected to the first seat, in the first round of tabulation (i.e., $t_{w_1,1} \geq \quota$), after any batch elimination has taken place.

\textbf{Prior Work:} In the approach of \cite{blom2022stvrlasSHORT}, an $\IQ$ assertion is formed to verify that $w_1$ has a quota on their first preferences, $\IQ(w_1)$.  
They then establish an estimated upper bound, $\overline{\tau}_{w_1}$, on the transfer value of ballots from $w_1$ using an assertion of the form $\UT(w_1, \overline{\tau}_{w_1})$.
Using this upper bound they create assertions to show that $w_2$ will always have a higher tally than all other candidates using \textsf{NL} assertions.  The method continues to increase the transfer value upper bound $\overline{\tau}_{w_1}$, until the sample size of the resulting audit increases, or $\overline{\tau}_{w_1}$ reaches 2/3 (the maximum transfer value in a 2 seat STV election).

\textbf{New Approach:} We vary this approach by introducing additional types of assertions to reduce the expected sample sizes required in an audit. The assertions (beyond \textsf{AG}) we use are (starred assertions are new):

\begin{description}
\item[$\IQ(c)$] Verifies that candidate $c$'s first-preference tally is equal to or greater than a quota: $t_{c,1} \geq \mathcal{Q}$.
\item[$\UT(c, \overline{\tau}_c)$] 
Assumes that candidate $c$ has been elected on their first preferences, and verifies that the transfer value for $c$ is less than $\overline{\tau}_c$: $t_{c,1} < \quota / (1 -
\overline{\tau}_c)$.
\item[$\LT(c, \underline{\tau}_{c})$] 
Assumes that $c$ has been elected on their first preferences, and verifies that the transfer value for $c$ is greater than $\underline{\tau}_{c}$: $t_{c,1} > \quota / (1 - \underline{\tau}_{c})$.
\item[$\AGS(w, l, W, \bm{\underline{\tau}}, \bm{\overline{\tau}})$] An extension of the \textsf{AG} assertion~\cite{blom2022stvrlasSHORT}.
The assertion shows that candidate $w$ will always have higher tally than candidate $l$ in the context where the candidates in $W$ have already been elected to a seat with lower and upper bounds on their transfer values $\bm{\underline{\tau}}$ and $\bm{\overline{\tau}}$. 

\hspace*{0.5cm}
The assertion compares the minimum tally of candidate $w$ in this context, with the maximum tally of $l$. %\hspace*{0.5cm}
In the original $\AG$ assertion \cite{blom2022stvrlasSHORT}, the minimum tally of $w$ consists only of those ballots on which $w$ is ranked first. The $\AGS$ assertion retains this and adds further counts to this minimum tally by including contributions from some ballots where $w$ is not ranked first. Specifically, for all ballots $b$ where $\first(\proj_{\cand - W}(b)) = w$, we reduce them in value by taking a product of transfer value lower bounds for the candidates in $W$ that precede $w$ in its ranking, and add these to $w's$ minimum tally.\footnote{In the actual election, there is a scenario where candidate $w$ does not get these ballots in their vote count: if $w$ is our next winner after those in $W$. In such a case, $w$ will be a winner, and thus we can ignore it since the context where we use these assertions is precisely to show that $w$ is a winner.}

For each ballot $b \in \ballots$, we define its contribution to the minimum tally of $w$, and the maximum tally of $l$, as follows.

\[C^{\AGS}_{min}(b, w, W, \bm{\underline{\tau}}, \bm{\overline{\tau}}) = 
\left\{
\begin{array}{ll}
1  & \first(b) = w \\
\prod_{k \in W'} \underline{\tau}_{k}\,\, & \first(\proj_{\cand - W}(b)) = w \\
 & \quad \text{ and } W' = \{c \in W : c \text{ precedes } w \text{ in } b\} \\
0 & \text{otherwise}
\end{array}
\right.
\]
\[C^{\AGS}_{max}(b, l, W, \bm{\underline{\tau}}, \bm{\overline{\tau}}) = 
\left\{
\begin{array}{ll}
0 & l \text{ does not occur in } b \\
0 & w \text{ appears before } l \text{ in } b \\
\textsf{maxt}(b,l,W,\overline{\tau}) & \first(b) \in W  \\
1 & \text{otherwise} \\
\end{array}
\right.
\]
where $\textsf{maxt}(b,l,W,\overline{\tau}) = \max\{\bm{\overline{\tau}}_c : c \in W \text{~precedes~} l \text{~in~}
b\}$. We define the minimum tally of $w$, $t1^{min}_w$, and the maximum tally of $l$, $t1^{max}_l$, as follows:
\begin{flalign}
t1^{min}_w & = \sum_{b \in \ballots} C^{\AGS}_{min}(b, w, W, \bm{\underline{\tau}}, \bm{\overline{\tau}})  \\
t1^{max}_l & = \sum_{b \in \ballots} C^{\AGS}_{max}(b, l, W, \bm{\underline{\tau}}, \bm{\overline{\tau}})  
\end{flalign}
We say that $\AGS(w, l, W, \bm{\underline{\tau}}, \bm{\overline{\tau}})$ iff $t1^{min}_w > t1^{max}_l$.
Note $\AG(w,l) \equiv \AGS(w,l,\emptyset,\_,\_)$.

\item[$\CNEBS(w,$ $l,W,$ $\bm{\underline{\tau}},\bm{\overline{\tau}},G^*,O^*)$]
An extension of the \textsf{NL} assertion~\cite{blom2022stvrlasSHORT}.
Establishes that candidate $w$ will always have a higher tally than candidate $l$ under the assumptions that: (i) the candidates in $W$ \textit{have already been seated}, with lower and upper bounds on their transfer values $\bm{\underline{\tau}}$ and $\bm{\overline{\tau}}$; (ii) $G^*$ denotes the candidates $g \in \cand$ for which $\AGS(g, l, W, \bm{\underline{\tau}}, \bm{\overline{\tau}})$ holds; and (iii) $O^*$ the candidates $o \in \cand$ for which $\AGS(w, o, W, \bm{\underline{\tau}}, \bm{\overline{\tau}})$ holds.  This contrasts with the assumptions underlying the original $\CNEB$ assertion, which only assumes that the candidates $W$ are seated \textit{at some point}. The assertion compares the minimum tally of $w$, in this context, against the maximum tally of $l$.

%This assertion considers a context where candidates $W$ have been seated, and their surpluses distributed, with upper bounds on their transfer values given by $\bm{\overline{\tau}}$ and lower bounds on their transfer values given by $\bm{\underline{\tau}}$. 
\hspace*{0.5cm}
 We define $w$'s minimum tally at a point at which they could be eliminated, where it is assumed that  $O^*$ have been prior eliminated. This minimum tally includes all ballots $b$ where $\first(\proj_{\cand - O^*}(b)) = w$, at value 1, and all ballots $b$ where $\first(\proj_{\cand - W}(b)) = w$, at a reduced value. For the maximum tally of $l$, we include all ballots on which $l$ precedes $w$ in their ranking, or $l$ appears and $w$ does not, excluding those on which a candidate $g \in G^*$ precedes $l$. 
For each ballot $b \in \ballots$, we define its contribution to the minimum tally of $w$, and the maximum tally of $l$, as follows.

\[C^{\CNEBS}_{min}(b, w, W, \bm{\underline{\tau}}, \bm{\overline{\tau}}, O^*) = 
\left\{
\begin{array}{ll}
1  & \first(\proj_{\cand - O^*}(b)) = w \\
\prod_{k \in W'} \underline{\tau}_{k}\,\, & \first(\proj_{\cand - W}(b)) = w \\
& \, \text{ and } W' = \{c \in W : c \text{ precedes } w \text{ in } b\}  \\
0 & \text{otherwise}
\end{array}
\right.
\]
\[C^{\CNEBS}_{max}(b, l, W, \bm{\underline{\tau}}, \bm{\overline{\tau}}, G^*) = 
\left\{
\begin{array}{ll}
0 & l \text{ does not occur in } b \\
0 & w \text{ appears before } l \text{ in } b \\
0 & \text{a } g \in G^* \text{ appears before } l \text{ in }  b\\
\textsf{maxt}(b,l,W,\overline{\tau})& \first(b) \in W  \\
1 & \text{otherwise} \\
\end{array}
\right.
\]
We define the minimum tally of $w$, $t2^{min}_w$, and the maximum tally of $l$, $t2^{max}_l$, as follows:
\begin{flalign}
t2^{min}_w & = \sum_{b \in \ballots} C^{\CNEBS}_{min}(b, w, W, \bm{\underline{\tau}}, \bm{\overline{\tau}}, O^*)  \label{eqn:NLS_Min}\\
t2^{max}_l & = \sum_{b \in \ballots} C^{\CNEBS}_{max}(b, l, W, \bm{\underline{\tau}}, \bm{\overline{\tau}}, G^*)  \label{eqn:NLS_Max}
\end{flalign}
We say that $\CNEBS(w, l, W, \bm{\underline{\tau}}, \bm{\overline{\tau}}, G^*, O^*)$ iff $t2^{min}_w > t2^{max}_l$.

% MB: Below not true as original NL does not include G* and O* as arguments, but
% G and O.
%Note that $\CNEB(w,l,W, \bm{\overline{\tau}},G^*, O^*)
%\equiv \CNEBS(w, l, W, \overline{0}, \bm{\overline{\tau}}, G^*, O^*)$ where $\overline{0}$ is a vector of zeros of length $|W|$.
\end{description}

\autoref{alg:FRW} outlines the procedure used to generate the assertions $\mathcal{A}$ of our new RLA for two-seat STV elections satisfying the first-round winner criterion. The prior approach  \cite{blom2022stvrlasSHORT} involved a single loop in which an upper bound on the first winner transfer value was incremented, and a candidate audit formed for each of these potential values for this upper bound. The original $\AG$ assertions were computed prior to this loop, as they did not take into account upper or lower bounds on the first winner's transfer value. Our new approach involves two loops -- the outer loop (steps 5-34) over potential values for the lower bound on the first winner's transfer value, $\underline{\tau}_{w_1}$, and the inner loop (steps 10-29) over potential values for the upper bound on the first winner's transfer value, $\overline{\tau}_{w_1}$. For each candidate value of $\underline{\tau}_{w_1}$, the inner loop searches for a value for $\overline{\tau}_{w_1}$ that results in the cheapest audit. The outer loop searches for a value for $\underline{\tau}_{w_1}$ for which the inner loop yields the cheapest overall audit. As per Blom~\emph{et al}~\cite{blom2022stvrlasSHORT}, the first assertion we create is $\IQ(w_1)$ to verify that our first winner, $w_1$, does indeed achieve a quota on their first preferences (step 1). Where a group elimination has taken place, and our resulting election satisfies the first-round winner criterion, the $\IQ(w_1)$ assertion verifies that $w_1$ has a quota on the basis of their first-preference tally \textit{and} any votes distributed to them from the group eliminated candidates.  

$\AGS$ assertions, which are used to help us form the $\CNEBS$ assertions required to show that $w_2$ beats all of the original losers, are formed inside the inner loop (step 14), allowing us to take advantage of both lower and upper bounds on the first winner's transfer value. When forming each $\CNEBS$, we add an $\AGS$ to our audit only if it allows us to reduce the expected ASN of the $\CNEBS$ we are trying to form, and where the ASN of the $\CNEBS$ without the $\AGS$ is higher than that of the $\AGS$ itself. In this way, we do not add $\AGS$ assertions to our audit where their benefit, in terms of making a $\CNEBS$ easier to audit, is outweighed by their cost.

Note that if our final audit contains an $\AGS$ and $\CNEBS$ with the same winner and loser, we remove the $\CNEBS$ from our audit as it is redundant.  

\begin{example}
\setlength{\parskip}{0mm}
    Consider again the 2021 Board and Estimates and Taxation election
    (Minneapolis, Minnesota).  \autoref{ex:2021BoEBatchElim} presents the first stage of an RLA for this election, identifying the assertions required to check that the two batch eliminated candidates did not have a mathematical possibility of winning. After the distribution of these eliminated candidates' ballots, we have a three candidate election that satisfies the first winner criterion. 
    
    S. Brandt is elected at this stage, with two remaining candidates (S. Pree-Stinson and P. Salica) vying for the second seat. Using the algorithm in \autoref{alg:FRW}, we form the assertion $\IQ(S. Brandt)$. The ASN for this assertion is 34 ballots.
    
    We then enter the outer loop at step 5 with a lower bound on S. Brandt's transfer value set to 0. (Where $\underline{\tau}_{w_1}$ is 0 we actually do not compute the associated $\LT$ assertion as it is not necessary).  Then, starting with an upper bound on S. Brandt's transfer value set to his actual transfer value plus $\delta$ (with $\delta = 0.05$), we enter the inner loop of the algorithm in \autoref{alg:FRW} at step 10.  The $\UT$ assertion required to show that S. Brandt's transfer value is less than, in this case, 0.3311, has an ASN of 131 ballots. $\AGS$ assertions are computed (step 14), and the $\CNEBS$ assertion required to show that S. Pree-Stinson never loses to P. Salica in the context where S. Brandt is seated first (steps 16 to 24). The $\CNEBS$ assertion has an ASN of 402 ballots. (In this case, none of the $\AGS$ assertions were found to be helpful in reducing the margin of this $\CNEBS$ assertion, and $\mathbf{AG'} \leftarrow \emptyset$ in step 22). At this stage, we have an RLA for the election that costs 402 ballots. The inner loop is repeated, with the upper bound on S. Brandt's transfer value set to 0.3811. The required $\UT$ assertion now costs 60 ballots. However, when proceeding to create the $\CNEBS$ assertion required to show that S. Pree-Stinson never loses to P. Salica, the assertion now costs 628 ballots. The new candidate configuration for our RLA is more costly, at 628 ballots, than the previous one, at 402 ballots. So, we break out of our inner loop at step 29. 

    We repeat our outer loop, with the lower bound on S. Brandt's transfer value now 0.1406 (or half of his actual transfer value, as per step 33). The $\LT$ assertion required to show that his transfer value is greater than this lower bound has an ASN of 59 ballots. We enter the inner loop at step 5 with the upper bound on S. Brandt's transfer value again set to 0.3311. The ASN of the $\UT$ assertion for this bound is, as before, 131 ballots. After proceeding through steps 14--24, we form an $\CNEBS$ assertion to show that S. Pree-Stinson never loses to P. Salica that now costs 247 ballots (again, we opt not to make use of any computed $\AGS$ assertions). We now have a  configuration for our audit that costs 247 ballots in total. When incrementing $\overline{\tau}$ for S. Brant to 0.3811, we do not improve upon this ASN (in fact, it will increase to 285). We break out of the inner loop at step 29.

    The outer loop will be repeated with $\underline{\tau}$ for S. Brandt increased to 0.1906. The required $\LT$ assertion for this bound will cost 87 ballots. By working through the inner loop, as before, we are able to find an audit configuration costing 217 ballots. Repeating the outer loop again with $\underline{\tau}$ for S. Brandt increased to 0.2406 gives us an audit  costing 194 ballots. The $\LT$ assertion will cost 184 ballots in this audit, the $\UT$ 131 ballots, and the required $\CNEBS$ 194 ballots. Incrementing $\underline{\tau}$ for S. Brandt again to 0.2906, in step 33, puts us beyond his actual transfer value of 0.2811. The outer loop condition fails, and we finish with an audit costing 194 ballots. This audit contains the assertions listed in \autoref{tab:2021BoERest}.  \qed

    \begin{table}[t]
\caption{Assertions verifying the election of S. Brandt and S. Pree-Stinson in the 2021 BoE election in Minneapolis, Minnesota, and their sample sizes.  }\centering
\begin{tabular}{ll|ll}
\hline
Assertion & ASN &  Assertion &  ASN \\
\hline
\multicolumn{2}{l|}{Batch elimination} & \multicolumn{2}{l}{Election of S. Brandt and S. Pree-Stinson} \\
\hline
    \AG(S. Brandt, UWIs) &  20  &   \IQ(S. Brandt) &  34 \\
    \AG(S. Pree-Stinson, UWIs) &  35 &  $\LT$(S. Brandt, 0.2406) &  184\\
    \AG(S. Brandt, K. Nikiforakis) &  27  &  \UT(S. Brandt, 0.3311) & 131\\
    \AG(S. Pree-Stinson, K. Nikiforakis) &  69  &  $\CNEBS$(S. Brandt, P. Salica, $\ldots$) & 194\\
\hline

Total cost: &  & & 194 \\
\hline
\end{tabular}
\label{tab:2021BoERest}
\end{table}

\end{example}

\begin{figure}[htbp]
\begin{tabbing}
xxx \= xxx \= xxx \= xxx \= xxx \= xxx \= xxx \kill
1  \> $iq \leftarrow \IQ(w_1)$ $\triangleright$ Form assertion to verify that $w_1$ has a quota on first preferences \\
2 \> $\underline{\tau}_{w_1} \leftarrow 0$ $\triangleright$ Lower bound on transfer value for first winner $w_1$\\
3 \> $ASN \leftarrow \infty$ $\triangleright$ $ASN$ of our audit\\
4 \> $\mathcal{A} \leftarrow \emptyset$ $\triangleright$ Assertions in our audit\\
5 \> {\bf while} $\underline{\tau}_{w_1} < \tau_{w_1}$ {\bf do } $\triangleright$ $\tau_{w_1}$ is the reported transfer value for $w_1$ \\
6 \> \> $lt \leftarrow \LT(w_1, \underline{\tau}_{w_1})$ $\triangleright$ Form $\LT$ assertion, denoted $lt$.\\
7 \> \> $\overline{\tau}_{w_1} \leftarrow \tau_{w_1} + \delta$ $\triangleright$ Upper bound on transfer value for first winner $w_1$\\
8 \>\> $\mathcal{A}' \leftarrow \emptyset$ \\
9 \>\> $ASN' \leftarrow ASN$ \\
10 \>\> \bf{while} $\overline{\tau}_{w_1} < 2/3$ \bf{do}\\
11 \>\>\>  $ut \leftarrow \UT(w_1, \overline{\tau}_{w_1})$ $\triangleright$ Form $\UT$ assertion, denoted $ut$.\\
12 \>\>\> $\mathcal{A}'' \leftarrow \{lt, ut, iq\}$ \\
13 \>\>\>  $ASN'' \leftarrow \max (lt.ASN, ut.ASN, iq.ASN)$ \\
\\
 \>\>\> $\triangleright$ Compute $\AGS$ assertions between $w_2$ and each $l \in losers$, and  \\
\>\>\>\> between each $l, l' \in losers$ such that $l \neq l'$ \\
14 \>\>\> $\bm{AG} \leftarrow $ [$\AGS(c, l, [w_1], [\underline{\tau}_{w_1}], [\overline{\tau}_{w_1}])\, | \,\forall c \in losers \cup \{w_2\}, l \in losers, c \neq l]$ \\
15 \>\>\> $O^*$ $\leftarrow$ $[c \,|\, \AGS(w_2, c, [w_1], [\underline{\tau}_{w_1}], [\overline{\tau}_{w_1}]) \in \bm{AG}]$\\
\\
\>\>\> $\triangleright$ Find $\CNEBS$ assertions to show that $w_2$ never loses to each $l \in losers$ \\
16 \>\>\> \textbf{for each } $l \in losers$ \textbf{do}\\
17 \>\>\>\> $G^*$ $\leftarrow$ $[c \,|\, \AGS(c, l, [w_1], [\underline{\tau}_{w_1}], [\overline{\tau}_{w_1}]) \in \bm{AG}]$\\
\\
18 \>\>\>\> $t2^{min}_{w_2} \leftarrow \sum_{b \in \ballots} C^{\CNEBS}_{min}(b, w_2, [w_1], [\underline{\tau}_{w_1}], [\overline{\tau}_{w_1}], O^*)$ (\autoref{eqn:NLS_Min}) \\
19 \>\>\>\> $t2^{max}_{l} \leftarrow \sum_{b \in \ballots} C^{\CNEBS}_{max}(b, l, [w_1], [\underline{\tau}_{w_1}], [\overline{\tau}_{w_1}], G^*)$
    (\autoref{eqn:NLS_Max}) \\
\\
20 \>\>\>\> \textbf{if} $t2^{min}_{w_2} > t2^{max}_{l}$ \textbf{then} \\
21 \>\>\>\>\> $nl \leftarrow \CNEBS(w_2, l, [w_1], [\underline{\tau}_{w_1}], [\overline{\tau}_{w_1}], G^*, O^*)$ \\
22 \>\>\>\>\> $\bm{AG}'$ $\leftarrow$ $\AGS$ assertions between $w_2$ and $o \in O^*$, and between \\
\>\>\>\>\>\> $g \in G^*$ and $l$, that were \textit{used to reduce} the ASN of $nl$ \\
23 \>\>\>\>\> $\mathcal{A}'' \leftarrow \mathcal{A}'' \cup \{nl\} \cup \bm{AG}'$\\
24 \>\>\>\>\> $ASN'' \leftarrow \max(ASN'', nl.ASN, a.ASN ~\forall a \in \bm{AG}')$ \\
\\
25 \>\>\>\textbf{if} $ASN'' < ASN'$ \textbf{then} \\
26 \>\>\>\> $\mathcal{A}' \leftarrow \mathcal{A}''$ \\
27 \>\>\>\> $ASN' \leftarrow ASN''$ \\
28 \>\>\>\> $\overline{\tau}_{w_1} \leftarrow \overline{\tau}_{w_1} + \delta$ \\
29 \>\>\> \textbf{else} break\\
30\>\>\textbf{if} $ASN' < ASN$ \textbf{then} \\
31\>\>\> $\mathcal{A} \leftarrow \mathcal{A}'$ \\
32\>\>\> $ASN \leftarrow ASN'$ \\
33\>\>\> $\underline{\tau}_{w_1} \leftarrow \underline{\tau}_{w_1} + \delta$ if $\underline{\tau}_{w_1} > 0$ and $\frac{\tau_{w_1}}{2}$ otherwise\\
34 \>\> \textbf{else} break
\end{tabbing}
\caption{Algorithm for generating assertions $\mathcal{A}$ for the revised RLA
    of a two-seat STV election satisfying the first-round winner criterion. The
    two reported winners of the election are $w_1$ and $w_2$, and $losers$
    denotes the remaining candidates.  Given an assertion, $a$, we use the
    notation ``$a.ASN$'' to denote its ASN.}
\label{alg:FRW}
\end{figure}

\subsection{Evaluation}\label{sec:Evaluation}

We contrast the expected cost (ASN) of our new two-seat STV RLA (for elections
satisfying the first-round winner criterion) relative to the existing method
\cite{blom2022stvrlasSHORT}. For sample size estimations, we use a risk limit
of 10\%,  an expected error rate of 2 overstatements per 1000 ballots, and the
ALPHA risk function of SHANGRLA \cite{shangrlaSHORT}. For the algorithm shown
in \autoref{alg:FRW}, we use a value of 0.05 for $\delta$.
\autoref{tab:FWC_SelectedResults} contrasts the expected sample sizes required by
our 2-seat STV RLAs across a set of real 2-seat STV instances--four BoE
elections held in Minneapolis, Minnesota between 2009 and 2021, and four
elections held as part of the Australian Senate election in 2016 and 2019--and
a series of US and Australian (NSW) IRV elections re-imagined as 2-seat STV
contests. All these instances satisfy the first-round winner criterion.
Selected instances from the full set of 92 NSW Legislative Assembly (NSW-LA)
elections, and 23 US IRV elections, are shown in
\autoref{tab:FWC_SelectedResults}.

Across the full set of 92 NSW-LA elections (re-imagined as 2-seat STV), no RLA could be formed using the prior approach \cite{blom2022stvrlasSHORT} for 8 instances. With the new method, four of these instances become auditable--although in one case, Lismore, the cost is still quite high at 2180 ballots. Across the remaining 84 instances, the new approach reduces required sample sizes by 15\% on average. Across the full set of 23 US IRV elections (re-imagined as 2-seat STV), no RLA could be formed for six instances using the prior approach \cite{blom2022stvrlasSHORT}. Using the new approach, three of these instances become auditable, with sample sizes of 3925, 350, and 166, as shown in \autoref{tab:FWC_SelectedResults}. Again, the new method reduces sample sizes for the remaining 17 elections by 15\% on average. For the Australian Senate and Minneapolis STV elections, the new method reduces required sample sizes by 19\%, on average.

\begin{table}[htbp]
\caption{ASNs for our 2 seat STV RLAs, comparing the original method of \cite{blom2022stvrlasSHORT} against the revised method. %We consider four elections held in Minneapolis, Minnesota (2009-21), four held as part of  the Australian Senate election (2016-1), and a series of US and Australian IRV elections re-imagined as 2 seat STV contests.  
All instances satisfy the first-round winner criterion. Where the revised method improves on the original, ASNs are in bold.}
\centering
\begin{tabular}{p{100pt}p{25pt}p{40pt}p{40pt}cc}
\toprule
          &            &                &                &    \multicolumn{2}{c}{ASN} \\
\cmidrule{5-6}
 Instance & $|\cand|$ & $|\mathcal{B}|$ & $\mathcal{Q}$  &    Prior RLA  &  New RLA \\
 \midrule
MN BoE 2009 & 7 & 32086 & 10696 & 191 & {\bf 100}\\
MN BoE 2013 & 5 & 48855 & 16286 & 33 & {\bf 31} \\
MN BoE 2017 & 4 & 69694 & 23232 & 23 & 23 \\
MN BoE 2021 & 5 & 95625 & 31876 & 402 & {\bf 194}\\
\midrule
AU Senate'16 ACT & 22 & 254767 & 84923 &  77 & {\bf 58} \\
AU Senate'19 ACT & 17 & 270231 & 90078 & 131 & {\bf 98}\\
AU Senate'16 NT &  19 &  102027 & 34010 & 60  & 60\\
AU Senate'19 NT &  18 & 105027 & 35010 &  58 & 58\\
\midrule
\multicolumn{5}{l}{US IRV elections re-imagined as 2 seat STV} \\
\midrule
MN Mayor 2013 & 36 & 79415 & 26472 & 73 & 73\\
Aspen'09 Mayor & 5 & 2528 & 843 & 43 & {\bf 41} \\
Berkeley'10 D1 CC & 5 & 5700 & 1901 & 60 & {\bf 29}\\
%Berkeley'10 D4 CC & 5 & 4759 & 1587 & 33 & 23 \\
Oakland'10  D4 CC & 8 & 20994 & 6999 & 82 & {\bf 64} \\
Oakland'10 Mayor & 11 & 119607 & 39870 & -- & {\bf 3925} \\
Oakland'10 D6 CC & 4 & 12911 & 4304 & -- & {\bf 350} \\
% Pierce'08 CA & 4 & 153528 & 51177 & 16 & 16 \\
Pierce'08 CE & 5 & 299132 & 99711 & -- & {\bf 166} \\
%San Fran'07 & 13 & 144730 & 48244 & -- & -- \\
%San Leandro'10 D5 CC & 7 & 22484 & 7495 & 106 & 106 \\
\midrule
\multicolumn{5}{l}{NSW'19 Legislative Assembly elections re-imagined as 2 seat STV} \\
\midrule
%Albury & 5 & 47632 & 15878 & 34 & {\bf 31}\\
%Auburn & 5 & 44842 & 14948 & 23 & 22\\
Ballina & 6 & 50127 & 16710 & 66 & 66\\
%Balmain & 6 & 49299 & 16434 & 57 & {\bf 43}\\
%Bankstown& 4 & 43730 & 14577 & 27 & 24\\
Bathurst&6 & 50833 & 16945 & 81 & {\bf 57}\\
%Baulkham Hills & 6 & 50310 & 16771 & 35  & 30\\
%Bega & 6 & 50701 & 16901 & 24 &  22\\
%Blacktown & 5 & 46909 & 15637 & 34 & 30 \\
%Blue Mountains & 7 & 49228 & 16410 & 65  & 62\\ 
%Cabramatta & 5 & 47343 & 15782 & 43 & 35\\
%Camden& 7 & 62772 & 20925 & 27  & 27\\
%Campbelltown&7 & 45526 & 15176 & 42 & 34 \\
%Canterbury&4 & 48756 & 16253 & 39 &  36\\
%Castle Hill&4 & 52571 & 17524 & 52 & 38\\
%Cessnock& 5 & 49781 & 16594 & 47 &  41\\
%Charlestown& 4 & 48874 & 16292 & 43 &  37\\
Clarence & 6 & 49355 & 16452 & 147 &  {\bf 84}\\
Coffs Harbour & 8 & 47333 & 15778 & 1225 & {\bf 515}\\
%Coogee & 8 & 46035 & 15346 & 29 &  29\\
Cootamundra & 6 & 47448 & 15817 & -- & --\\ 
%Cronulla & 5 & 51141 & 17048 & 34 & 31 \\
%Davidson & 5 & 49169 & 16390 &  -- & --\\
%Drummoyne & 5 & 47464 & 15822 & 35 & 30\\ 
%Dubbo & 7 & 48455 & 16152 & 56 & 56\\
%East Hills & 7 & 47623 & 15875 & 26 & 26\\ 
%Epping & 5 & 49224 & 16409 & 30 & 27\\
%Fairfield & 4 & 44075& 14692 & 51 & 42\\ 
%Gosford & 6 & 48637 &16213 & 24 & 23\\
%Goulburn & 7 & 51057 & 17020 &39 & 39\\
%Granville & 8 & 44191 & 14731 & 17 & 16\\
%Hawkesbury & 9 & 48941 & 16314 & 61 & 51\\
%Heathcote & 5 & 51334 & 17112 & 18 & 18\\
Heffron & 5 & 50010 & 16671 & 1778 & {\bf 211}\\
Holsworthy & 6 & 48244 & 16082 & 20 & 20\\
%Hornsby & 9 & 50003 & 16668 & 105 & {\bf 78}\\
%Keira & 4 & 51936 & 17313 & 116 & {\bf 71}\\
%Kiama & 5 & 48946 & 16316 & 33 & 29\\
%Kogarah & 5 & 45576 & 15193 & 25 & 25\\
Ku-ring-gai & 6 & 48730 & 16244 & 202 & {\bf 99}\\
Lake Macquarie & 6 & 50082 & 16695 & 129 & {\bf 73}\\
%Lakemba & 6 & 44615 & 14872 & 52 & 43\\
Lane Cove & 6 & 50941 & 16981 & 132 & {\bf 109} \\
Lismore & 7 & 48145 & 16049 &  -- &  {\bf 2180}\\
%Liverpool & 6 & 47035 & 15679 & 29 & 27\\
%Londonderry & 5 & 52686 & 17563 & 15 & 15\\
%Macquarie Fields & 6 & 52789 & 17597 & 26 & 24\\
%Maitland & 8 & 53050 & 17684 & 36 & 34\\
Manly & 6 & 48316 & 16106 & 363 & {\bf 150}\\
%Maroubra & 7 & 48278 & 16093 & 36 & 35\\
%Miranda & 6 & 49066 & 16356 & 27 & 25\\
%Monaro & 6 & 49448 & 16483 & 28 & 26\\
%Mount Druitt & 5 & 47031 & 15678 & 28 & 24\\ 
%Mulgoa & 4 & 51375 & 17126 & 17 & 16\\
%Murray & 10 & 47233 & 15745 & 42 & 42\\
%Myall Lakes & 6 & 50315 & 16772 & 24 & 24\\
Newcastle & 8 & 50319 & 16774 & 214 & {\bf 173}\\
%Newtown & 7 & 46312 &15438 & 46 & 31\\
North Shore & 9 & 47774 & 15925 & 470 & {\bf 184}\\
Northern Tablelands & 4 & 48678 & 16227 & -- & {\bf 143} \\
% Oatley & 5 & 48120 & 16041 & 18 & 17\\
%Orange & 9 & 50295 & 16766 & 36 & 33 \\
Oxley & 5 & 48540 & 16181 & 98 & {\bf 80}\\
%Parramatta & 7 & 48728 & 16243 & 22 & 21\\ 
%Penrith & 10 & 48853 & 16285 & 34 & 34\\
Pittwater & 8 & 49119& 16374 & 163 & {\bf 110}\\
%Port Macquarie & 4 & 52735 & 17579 & 39 & 31\\
% Port Stephens & 6 & 49642 & 16548 & 15 & 15\\
% Prospect & 5 & 46845 & 15616 & 24 & 22\\
% Riverstone & 3 & 53510 & 17837 & 14 & 14\\
%Rockdale & 6 & 47892 & 15965 & 22 & 21\\
%Ryde & 8 & 48492 & 16165 & 22 &  22\\
%Seven Hills & 6 & 46977 & 15660 & 16 & 16 \\
%Shellharbour & 4 & 54840 & 18281 & 42 & 34\\
%South Coast & 3 & 48880 & 16294 & 29 & 26\\
%Strathfield & 6 & 46217 & 15406 & 20 & 20\\
Summer Hill & 6 & 48785 & 16262 & -- & {\bf 110}\\
%Swansea & 5 & 49192 & 16398 & 22 & 21\\
%Sydney & 6 & 43201 & 14401 & 67 & 58\\
Tamworth & 6 & 50578 & 16860 & -- & {\bf 129}\\ 
%Terrigal & 7 & 50284 & 16762 & 33 & 31\\
%The Entrance & 7 & 48086 & 16029 & 23 &  23\\
%Tweed & 5 & 46661 & 15554 & 31 & 30\\
%Upper Hunter & 8 & 48525 & 16176 & 460 & 460\\ 
Vaucluse & 7 & 46023 & 15342 & -- & {\bf 325}\\
% Wagga Wagga & 7 & 48578 & 16193 & 82 & 59 \\
% Wakehurst & 7 & 48712 & 16238 & 126 & 92\\
Wallsend & 5 & 51351 & 17118 & 149 &  {\bf 83}\\
Willoughby & 8 & 47857 & 15953 & -- & -- \\
Wollondilly & 8& 50989 & 16997 & 120 & {\bf 88}\\
Wollongong & 7& 51435 & 17146 & 205 & {\bf 123} \\
%Wyong & 4 & 48121 & 16041 & 30 & 28 \\
\bottomrule
\end{tabular}
\label{tab:FWC_SelectedResults}
\end{table}

% ---------------------------------------------------------------------------

\section{Scenario: General Method}\label{sec:GeneralMethod}

The \textit{general method} of Blom~\emph{et al}~\cite{blom2022stvrlasSHORT} describes how we can form an RLA for a 2-seat STV election where no candidate has a quota on their first preferences. We do not, in this paper, present an improvement to this approach--in the sense of enabling audits for instances that we could not previously audit.   We do, however, show how we can adapt the method to perform \textit{partial} audits of elections where a full RLA, verifying both winners, is not possible. The experiments of Blom~\emph{et al}~\cite{blom2022stvrlasSHORT} demonstrate that forming full RLAs for this class of 2-seat STV elections is challenging, and generally not possible with existing methods.

A partial RLA can be used to verify \textit{some} aspects of the election outcome. For example, that some reported losers did indeed lose, and that one of the reported winners did indeed win. In this paper, we reframe the general method into five stages.  We still use the original assertion types, $\AG$ and $\CNEB$, as we are not assuming that one or more candidates have been \textit{previously} seated. Stages 1, 2, and 3 are present in the general method of Blom~\emph{et al}~\cite{blom2022stvrlasSHORT}. Stages 4 and 5 are introduced in this paper to (i) describe how we can form a partial RLA for a 2-seat STV election when a full RLA cannot be formed (Stage 5) and (ii) reduce the required sample size of the resulting partial or full RLA (Stage 4).

\begin{enumerate}
\setlength{\itemsep}{1mm}
\setlength{\parskip}{1mm}
    \item \textbf{Form $\AG$ Assertions} For each pair of candidates $c, c' \in \cand$, we determine whether we can form the assertion $\AG(c, c')$. We keep track of each $\AG$ that we can form, and its cost.

    \item \textbf{Rule out candidates (find Definite Losers)} We use the $\AG$ assertions that we formed in Stage 1 to determine whether some candidates definitely lost the election. All candidates $c \in \cand$ for which there exists \textit{at least two} other candidates $c', c'' \in \cand - \{c\}$ such that $\AG(c', c)$ and $\AG(c'', c)$ \textit{definitely lost} the election. We denote this set of candidates $DL$, and the set of $\AG$ assertions required to show that these candidates definitely lost as $\mathcal{A}_{DL}$. This set will contain two $\AG$ assertions for each definite loser. The maximum sample size required to audit any assertion in this set is denoted the Stage 2 sample size.

    \item \textbf{Rule out alternate winner pairs} We consider all pairs of candidates from the set $\cand - DL$, excluding the pair of reported winners,  as potential alternate winner outcomes. We follow the approach of Blom~\emph{et al}~\cite{blom2022stvrlasSHORT}, and attempt to rule out each of these alternate winner pairs with an $\CNEB$ assertion. For a pair $(c_1,c_2)$, we first assume that $c_1$ is seated \textit{at some point}, and look for another candidate $c'$ that never loses to $c_2$ in this context: 
    \begin{center}
    $\CNEB(c', c_2, [c_1], G, O)$
    \end{center}
    where $G$ is the set of candidates $g$ for which $\AG(g, c_2)$ and $O$ the candidates $o$ for which $\AG(c', o)$. We compare the cost of this $\CNEB$ assertion with one formed when we assume that $c_2$ is seated at some point, and look for a $c'$ that never loses to $c_1$:
    \begin{center}
    $\CNEB(c', c_1, [c_2], G, O)$ 
    \end{center}
    where $G$ is the set of candidates $g$ for which $\AG(g, c_1)$ and $O$ the candidates $o$ for which $\AG(c', o)$. The cheapest $\CNEB$, assuming we are able to form at least one, is used to rule out the outcome of $c_1$ and $c_2$ winning together. We denote the set of assertions used to rule out candidate pairs in this stage, $\mathcal{A}_3$. This set includes the formed $\CNEB$ assertions and any $\AG$ assertion used to reduce the margin of those $\CNEB$ assertions. The maximum sample size required to audit an assertion in this set the Stage 3 sample size.

    \item \textbf{Reduce audit sample size} Ruling out a candidate $c$ in Stage 2, by looking for two other candidates who tallies are always greater than $c$, may be unnecessarily costly. We may have been able to rule out all alternate winner pairs involving $c$ with cheaper $\CNEB$ assertions in Stage 3. 
    
%We sort the candidates in $DL$ in terms of the sample size required to rule them out in Stage 2. 
While the Stage 2 sample size is higher than that of Stage 3, we take the current `most difficult to rule out' candidate in $DL$, $d$. Let $ASN^2_d$ denote the sample size required to rule out $d$ as a potential winner in Stage 2. We form a set of alternate winner pairs by pairing $d$ with all candidates in $\cand - DL$. We perform the Stage 3 process over this new pair set. 
    If the sample size required to rule out these pairs, $ASN^3_d$, is less than $ASN^2_d$, we: 
    \begin{enumerate}
    \item Remove the assertions formed in Stage 2 to rule out $d$ from  $\mathcal{A}_{DL}$; 
    \item Add the new assertions formed to rule out all alternate winner pairs involving $d$ to $\mathcal{A}_3$; 
    \item Update the Stage 2 sample size, excluding the cost of ruling out $d$; 
    \item Update the Stage 3 sample size to include $ASN^3_d$; and,
    \item Remove $d$ from $DL$. 
    \end{enumerate}
    
    If $ASN^3_d \geq ASN^2_d$, or we could not rule out the new set of alternate winner pairs, we do not change our audit and move to Stage 5. Otherwise, we take the next most difficult to rule out candidate in $DL$, and repeat Stage 4.

    \item \textbf{Summarise what can (and cannot) be audited} If we have been able to rule out each alternate winner pair with an $\CNEB$, we have a full RLA. This RLA contains the assertions in $\mathcal{A}_{DL}$ and $\mathcal{A}_3$. If we were not able to rule out every alternate winner pair, we consider whether the ones we could rule out imply that some additional candidates definitely lost or definitely won. Let $Rem$ denote the set of alternate winner pairs that we could not rule out. 
    \begin{itemize}
        \item \textbf{Definite Winners} If there is a candidate $c$ present in \textit{every} remaining pair in $Rem$, we add this candidate to a set $DW$.     
        \item \textbf{Definite Losers} Includes all candidates in $DL$ (Stage 2) and any  $c \in \cand - DL$ (excluding the reported winners) that is not present in any of the alternate winner pairs in $Rem$. Each such $c$  is added to $DL$.
        \item \textbf{Potential Winners} All candidates in the set $\cand - DL$ are potential winners. These are reported losers and winners whose elimination or election we could not verify. 
    \end{itemize}
    Our partial RLA contains the assertions in the set $\mathcal{A}_{DL} \cup \mathcal{A}_3$ and can be used to establish that the candidates in the set $DL$ definitely lost, and that the candidates in the set $DW$ definitely won.
\end{enumerate}

%\mb{Possibly instead of example 4, we highlight, like we do in the paragraph after example 4, how often we can show that at least one of the winners definitely won, how many candidate percentage wise definitely lost etc.}
\begin{example}
    Consider the 2022 Australian Senate election for ACT, a 2-seat STV election that does not satisfy the first-round winner criterion. The quota for this election was 95073.  None of the 23 candidates have a quota on their first preferences. GALLAGHER won her seat after the elimination of 11 candidates. After a further 9 candidates were eliminated, POCOCK won the second seat. In Stage 1, we can form 44  $\AG$ assertions with ASNs ranging from 14 to 1141. In Stage 2, we use some of these $\AG$s to mark 16 candidates as \textit{definite losers}. The maximum ASN of the assertions we use in this stage is 775. In Stage 3, there are 20 alternate winner pairs that we can form with the remaining 7 candidates. We are able to rule out all but three of these with $\CNEB$ assertions, requiring a sample size of 420 ballots. The most expensive candidate to rule out as a winner in Stage 2  requires a sample of 775 ballots. In Stage 4, we take this `expensive to rule out candidate', $d$, and rule out all alternate winner pairs involving $d$ with $\CNEB$ assertions instead. These $\CNEB$ assertions have a maximum ASN of 47 ballots. Our overall audit cost reduces to 420 ballots. Our Stage 5 summary indicates that we can show that GALLAGHER correctly won, and that four of the remaining 22 candidates, including POCOCK, are potential winners.
\qed
\end{example}

We have collected data for 587 3-4 seat STV elections taking place in 2017 and 2022 to elect local councillors in Scotland. These elections involve 3 to 13 candidates. When viewed as 2-seat STV elections, 428 of these instances involve an elimination in the first round. A full RLA can be formed for 68 of these 428 instances. For this set of 68 instances, performing Stage 4 reduces required sample sizes by 58\% on average for 20 elections of the 68 (reductions range from 5\% to 97\%), and makes no difference on required sample size for the remaining 48. For the 360/428 instances for which a full RLA could not be formed, performing Stage 4 reduces the required sample size of the partial RLA in 122/360 of these instances (by 49\% on average, from 2\% to 99\%).

% ---------------------------------------------------------------------------

\section{Concluding Remarks}\label{sec:conc}

Auditing STV elections is a challenging problem, but one of very real interest given the common use of STV throughout the world. 
The main challenge arises as ballots can change their value across tabulation.  In this paper we have shown how reasoning about both lower and upper bounds on transfer values may improve our ability to audit 2-seat STV elections.
The revisions substantially reduce the number of ballots expected to be required to audit an election, and in some cases makes it possible to audit an election that the previous method~\cite{blom2022stvrlasSHORT} could not. 
We also show how to effectively audit batch elimination, as well as partially audit elections where no candidate gets a quota initially. 
While significant advances are still required to get to the point of auditing large Australian Senate elections, STV elections with 6 seats and over 100 candidates and say 4 million ballots, the new techniques we develop here help us on the path to this goal.

% ---- Bibliography ----
%
% BibTeX users should specify bibliography style 'splncs04'.
% References will then be sorted and formatted in the correct style.
%
\vspace*{-4mm}
\bibliographystyle{splncs04}
\bibliography{BIB}

% ---------------------------------------------------------------------------

\end{document}